\documentclass[smus]{snow2e}
\usepackage{epsf}

\begin{document}

\title{Discovery Mass Reach for Excited Quarks at Hadron Colliders}

\author{Robert M. Harris\\ {\it Fermilab, Batavia, IL  60510}}

\maketitle

\thispagestyle{empty}\pagestyle{empty}

\begin{abstract} 
%
If quarks are composite particles then excited states are expected. 
We estimate the discovery mass reach as a function of integrated luminosity 
for excited quarks decaying to dijets at the Tevatron, LHC, and a Very Large 
Hadron Collider (VLHC). At the Tevatron the mass reach is $0.94$ TeV for Run II 
(2 fb$^{-1}$) and $1.1$ TeV for TeV33 (30 fb$^{-1}$). At the LHC the mass
reach is $6.3$ TeV for 100 fb$^{-1}$. At a VLHC with a center of mass energy
,$\sqrt{s}$, of 50 TeV (200 TeV) the mass reach is 25 TeV (78 TeV) for an integrated 
luminosity of $10^4$ fb$^{-1}$. However, an excited quark with a mass of 
25 TeV would be discovered at a hadron collider with $\sqrt{s}=100$ TeV and an 
integrated luminosity of 13 fb $^{-1}$, illustrating a physics example 
where a factor of 2 in machine energy is worth a factor of $1000$ in luminosity.

\end{abstract}
\section{Excited Quarks}
We consider a model of composite quarks with excited states that have spin 1/2
and weak isospin 1/2. The effective Lagrangian for chromomagnetic transitions 
between excited quarks ($q^*$) of mass $M$ and common quarks (q) is 
constrained by gauge invariance to be~\cite{ref_qstar}:
\begin{equation}
\label{eq_Lagrangian}
{\cal L} = \frac{g_s f_s}{4M}\bar{q}^*_R \sigma^{\mu\nu} \lambda_a
G_{\mu\nu}^a q_L \ + \ h.c.
\end{equation}
where $G^a$ are gluon fields, $\lambda_a$ are SU(3) structure constants, 
and $g_s$ is the strong coupling. Here we have chose the compositeness
scale to be $\Lambda=M$, by writing $M$ in the denominator in 
Eq.~\ref{eq_Lagrangian}, because the excited quark mass should be close to the 
energy scale of quark compositeness. The constant $f_s$ depends on 
the unknown dynamics of the quark consituents, and is generally assumed to 
be equal to 1, thereby giving standard model couplings. Excited quarks decay 
to common quarks via the emission of a gluon in approximately 83\% of all 
decays. Excited quarks can also decay to common quarks by emitting a W, Z, or 
photon, through an effective Lagrangian similar to 
Eq.~\ref{eq_Lagrangian}. 

We consider the process $qg\rightarrow q^* \rightarrow qg$ for 
discovering an excited quark at a hadron collider.
The signal is two high energy jets, resulting from hadronization of the final 
state quark and gluon, which form a peak in the dijet invarian mass 
distribution. The subprocess differential cross section is a Breit-Wigner:
\begin{equation}
\frac{d\hat{\sigma}}{d\hat{t}} = \frac{2\pi\alpha_s^2}{9M^4}
\frac{\hat{s}}{ \left( \hat{s} - M^2\right)^2 + \Gamma^2 M^2}
\label{eq_xsec}
\end{equation}
where $\alpha_s$ is the strong coupling, $\hat{s}$ and $\hat{t}$ are 
subprocess Mandelstam variables, and $\Gamma$ is the width of the excited
quark. The sum of the partial widths in the gluon, W, Z, and photon channels, 
gives a half width at half maximum of $\Gamma/2 \approx 0.02M$. 

In Eq.~\ref{eq_xsec} we have already averaged over the angular 
distribution in the center of mass frame, $dN/d\cos\theta^* \sim 1 +
\cos\theta^*$, where $\theta^*$ is the angle between the initial state 
and final state quark in the subprocess center of mass frame.
In hadron collisions this subprocess angular distribution results in an 
isotropic dijet angular
distribtion $dN/d\cos\theta^* \sim 1$.  This is because for every quark in 
hadron 1 that becomes an excited quark and emerges in the final state at a
fixed $\cos\theta^*$, with rate proportonal to $1 + \cos\theta^*$, there is
a quark in hadron 2 which is headed in the opposite direction, and emerges at
the same value of $\cos\theta^*$ with rate proportional to $1 - \cos\theta^*$.
The sum of the two angular distributions is isotropic.

\section{Background and Cuts}

Normal parton-parton scattering via QCD produces a large background to the
dijet decays of excited quarks. 
However, QCD is dominated by t-channel gluon
exchange which gives a dijet angular distribution $dN/d\cos\theta^*\sim
1/(1-\cos\theta^*)^2$, where $\theta^*$ is the angle between the incoming
parton and the jet in subprocess center of mass. In contrast excited quark 
production and decay results in an isotropic dijet 
angular distribution as discussed above.  Therefore to 
suppress QCD
backgrounds we require $|\cos\theta^*|<2/3$ and we also require the
pseudorapidity of each jet satisfy $|\eta|<2$.  We note that any dijet 
analysis will 
generally make a $|\cos\theta^*|$ cut to have uniform trigger acceptance as
a function of dijet mass, and an $|\eta|$ cut is to stay within a defined
region of the detector.
We include all lowest order QCD subprocesses in our background calculation:
$qq\rightarrow qq$, $q\bar{q} \rightarrow gg$, $qg \rightarrow qg$, $gg
\rightarrow gg$ and $gg \rightarrow q\bar{q}$.

\section{Cross Section}
For both the excited quark signal and the lowest order QCD background, we 
convolute the subprocess differential cross section with CTEQ2L parton 
distributions~\cite{ref_cteq} of the colliding hadrons, within the above range of
$\cos\theta^*$ and $\eta$.
This gives the differential cross section as a function
of dijet mass, $d\sigma/dm$, for both the excited quark signal and the lowest
order QCD background.  For the excited quark signal we consider only the 
first generation, $u^*$ and $d^*$, and we assume they are degenerate in mass.
The half width of the excited quark resonance
remains $\Gamma/2 \approx 0.02M$.
This is significantly more narrow than the dijet mass 
resolution at the Tevatron,
which is roughly Gaussian with RMS 
\onecolumn
\begin{figure}[tbh]
\hspace*{-0.25in}
\epsffile{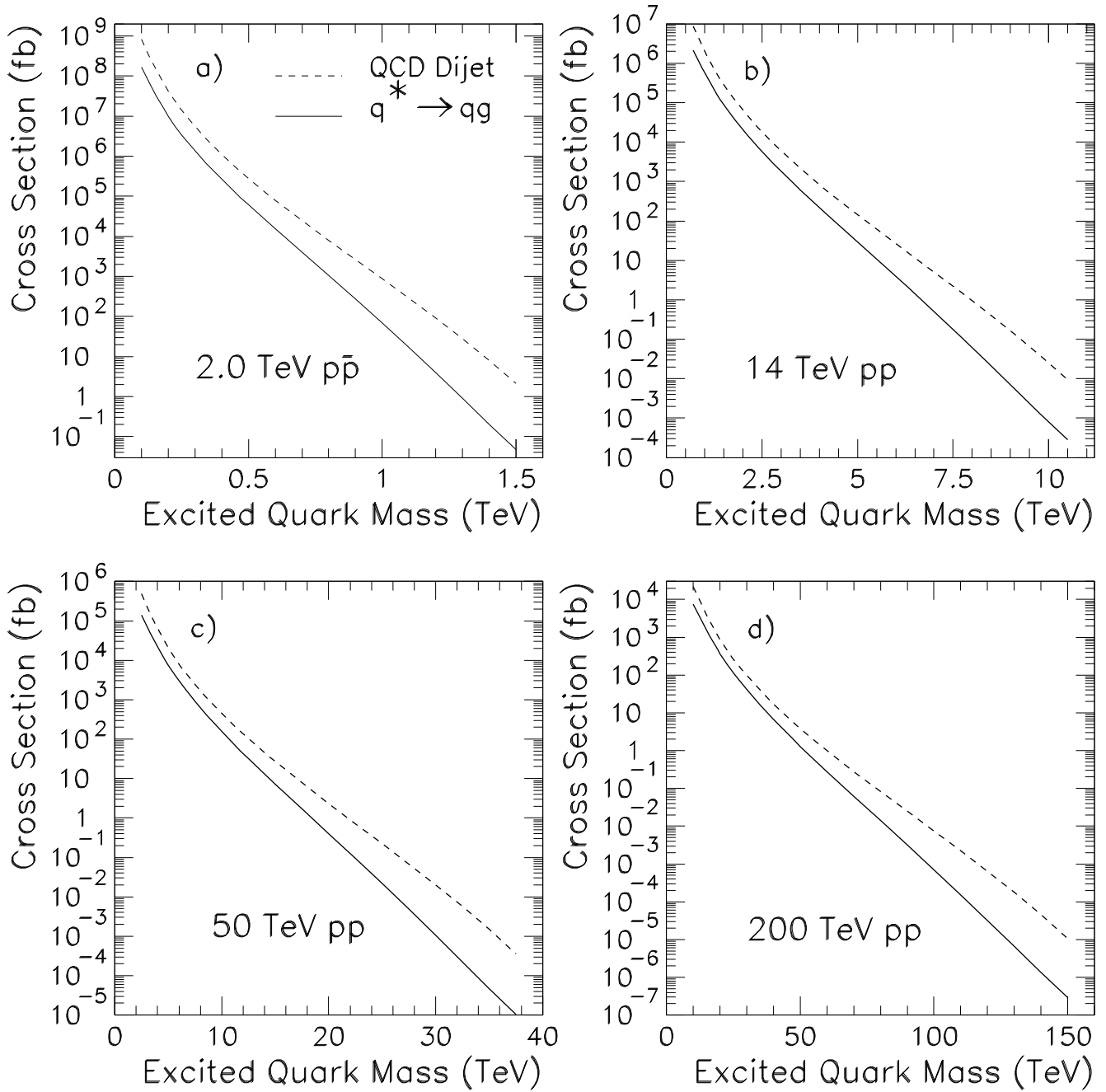}
\caption[]{ Lowest order parton level cross sections within a 16\% wide search
window for QCD dijets (dashed curve) and excited quarks decaying to dijets
(solid curve) are shown as a function of excited quark mass at {\bf a)} the 
future energy of the Tevatron, {\bf b)} the LHC, {\bf c)} a VLHC with center
of mass energy 50 TeV, and {\bf d)} 200 TeV.
All cross sections are for dijets with $|\eta|<2$, $|\cos\theta^*|<2/3$.}
\label{fig_xsec}
\end{figure}

\begin{figure}[tbh]
\hspace*{-0.25in}
\epsffile{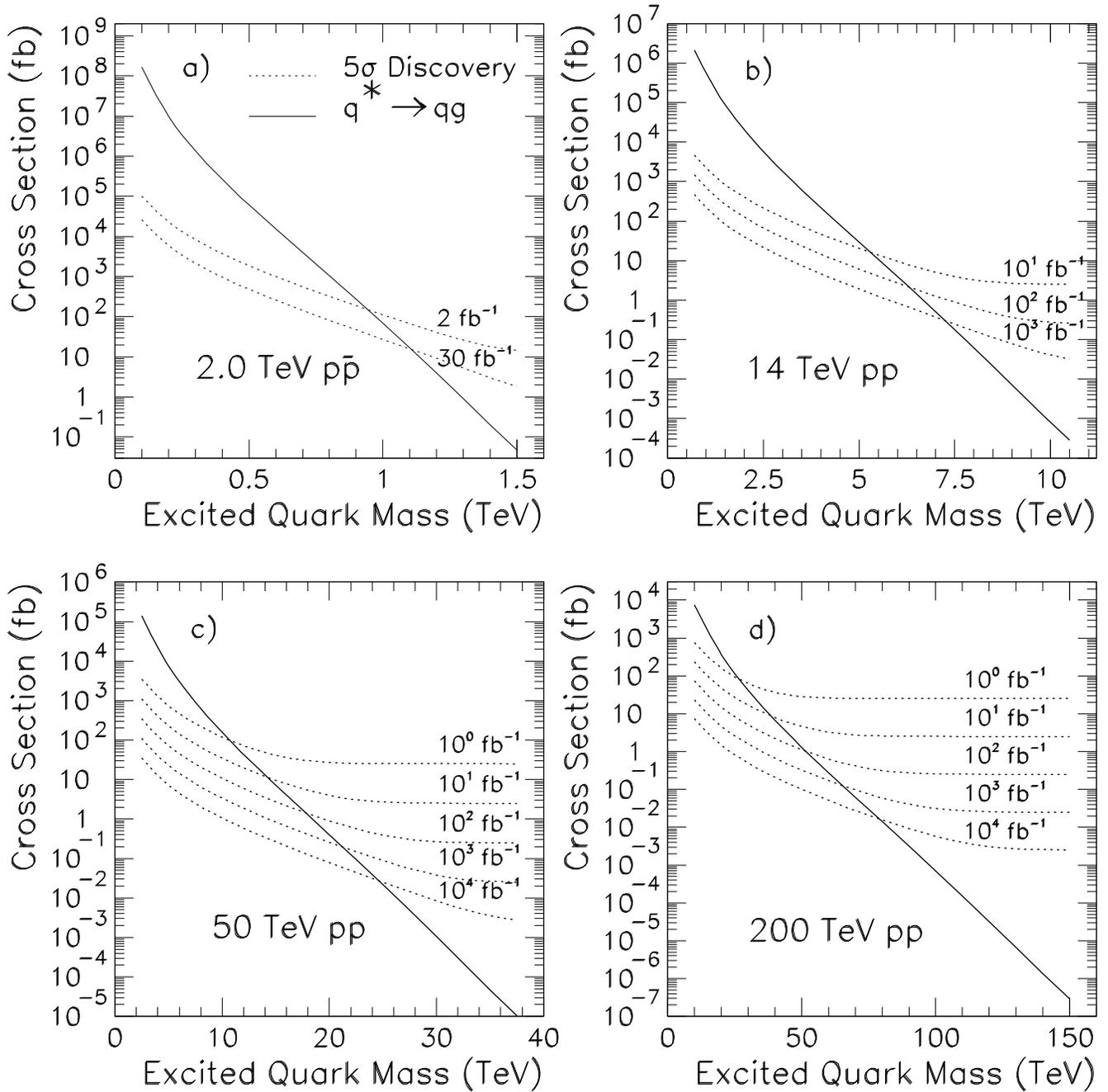}
\caption[]{ 
The predicted cross section for
dijet decays of excited quarks (solid curve) is compared to the 5$\sigma$ 
discovery reach (dotted curves) at various luminosities for 
{\bf a)} the 
future energy of the Tevatron, {\bf b)} the LHC, {\bf c)} a VLHC with center
of mass energy 50 TeV, and {\bf d)} 200 TeV. All
cross sections are for dijets with $|\eta|<2$, $|\cos\theta^*|<2/3$, and
invariant mass within 16\% of the excited quark peak assuming a 10\% dijet mass
resolution.}
\label{fig_reach}
\end{figure}

\twocolumn

\noindent
deviation $\sigma \approx 0.1M$. If we assume
a Gaussian dijet resolution of width $\sigma \approx 0.1M$ at all hadron 
colliders, then 90\% of the dijet events from an excited quark will be inside a 
16\% mass window $0.84M < m < 1.16M$, where $m$ is the dijet invariant mass. 
We integrate the differential cross section, $d\sigma/dm$, for 
both the excited quark signal and the QCD background within the 16\% mass
window to obtain an estimate of the signal and background cross section for a 
search. Figure~\ref{fig_xsec} shows the resulting total signal and background 
cross section in the search window at the Tevatron, LHC and VLHC as a function
of excited quark mass.

\section{Discovery Mass Reach}

The QCD background rate
is used to find the 5 $\sigma$ discovery cross section. This
is conservatively defined as the cross section which is
above the background by 5 $\sigma$, where $\sigma$ is the statistical error on 
the measured cross section (not the background).  For example, if the 
background were zero events the $5\sigma$ discovery rate would be 25 events.  
In Fig.~\ref{fig_reach} we compare the
excited quark cross section to the 5 $\sigma$ discovery cross section at 
various luminosities for the future Tevatron, the LHC, and the VLHC.
The excited quark discovery mass reach, defined as the mass at which an
excited quark would be discovered with a 5$\sigma$ signal, is tabulated as
a function of mass for the LHC and VLHC proton-proton colliders in
Table I.  We have also performed the calculation for VLHC proton-antiproton
colliders, where the QCD background is slightly higher but the excited quark
signal is exactly the same, which yields a 3\% smaller mass reach.
Because of space limitations, Figs.~\ref{fig_xsec} and \ref{fig_reach} do not
display curves for a 100 TeV VLHC, but the mass reach of a 100 TeV VLHC 
tabulated in Table I was determined from curves similar to those in 
Fig.~\ref{fig_reach}.

The mass reach at the future Tevatron is $0.94$ TeV for collider run II 
(2 fb$^{-1}$) and $1.1$ TeV for TeV33 (30 fb$^{-1}$).  This can be compared
to the published 95\% CL limit of 570 GeV from CDF~\cite{ref_qstar_cdf} and the
the preliminary limits of 750 GeV from CDF and 720 GeV from 
D0~\cite{ref_qstar_d0}.
The mass reach at the LHC is 6.3 TeV for 100 fb$^{-1}$, which could be 
obtained by running for one year ($\sim 10^7$ seconds) at the design
luminosity of $10^{34}$ cm$^{-2}$ s$^{-1}$.  Since the design luminosity may not
be quickly achieved, we note that with only 10 fb$^{-1}$ at the beginning of the
LHC the mass reach is still 5.3 TeV.  Ultimately, the LHC may be able to 
integrate 1000 fb$^{-1}$, which will provide a mass reach of 7.3 TeV.
The mass reach at the VLHC varies widely depending on the 
energy of the machine and it's luminosity.  A 50 TeV machine with only 1
fb$^{-1}$ of integrated luminosity has a mass reach of $10.5$ TeV,
significantly better than LHC with any conceivable luminosity. At the other
extreme, a 200 TeV machine with $10^4$ fb$^{-1}$ would have a mass 
reach of 78 TeV.  
\begin{table}[tbh]
Table I: The 5$\sigma$ discovery mass reach for excited quarks of a 
proton-proton collider
as a function of integrated luminosity is tabulated for the LHC with a 
center of mass energy of 14 TeV and the VLHC with a center of mass energy
of 50, 100 and 200 TeV. \\
\begin{center}
\begin{tabular}{|c|c|c|c|c|}\hline 
   & \multicolumn{4}{c|}{Excited Quark Mass Reach} \\ 
Integrated   & LHC & VLHC & VLHC & VLHC \\ 
Luminosity   & 14  &  50  &  100 & 200  \\ 
(fb$^{-1}$)  & (TeV)  & (TeV)   & (TeV)    & (TeV) \\ \hline
1 &  -- & 10.5 & 17 & 26 \\
10 & 5.3 & 14.0 & 24 & 39 \\
100 & 6.3 &  17.5 & 31 & 52 \\
$10^3$ & 7.3 & 21 & 38 & 65 \\
$10^4$ & -- &  24.5 & 45 & 78 \\ \hline
\end{tabular}
\end{center}
\end{table}

The mass reach in table I appears to be a smooth function of the proton-proton
center of mass energy, $\sqrt{s}$, and integrated luminosity, $L$.  The
following analytic function exactly reproduces the VLHC mass reach in Table I 
for the energy range $50<\sqrt{s}<200$ TeV and the luminosity
range $1<L<10^4$ fb$^{-1}$:
\begin{equation}
\label{eq_lum}
M = 7 + 3\log_2\left(\frac{\sqrt{s}}{50}\right) + k(1+\log_{10}L)
\end{equation}
where $k$ depends on the energy of the machine according to 
\begin{equation}
\label{eq_lum2}
k = \frac{7}{2} + \frac{11}{3}\left(\frac{\sqrt{s}}{50} -1\right) 
-\frac{1}{6}\left(\frac{\sqrt{s}}{50} - 1\right)^2
\end{equation}
Although Eq.~\ref{eq_lum} and \ref{eq_lum2} reproduces the VLHC mass reach,
at LHC these equations give a mass reach that is 40\% lower than the 
numbers in Table I. We provide Eq.~\ref{eq_lum} and \ref{eq_lum2} 
for interpolation among the VLHC entries in Table I only. We do not 
recommend these equations be used for extrapolation outside the energy
range $50<\sqrt{s}<200$ TeV and the luminosity range $1<L<10^4$ fb$^{-1}$.

\section{Energy vs. Luminosity}
To clarify the superior gains obtained by increasing the
energy of a machine, as opposed to increasing the luminosity, we show in 
Fig.~\ref{fig_lum} the mass reach for the VLHC which is also tabulated in
Table I.  Note that the mass reach is proportional to the logarithm of the 
luminosity, but is almost directly proportional to the energy of the machine.
To clarify the energy vs. luminosity tradeoff consider the following 
hypothetical case.

\subsection{Discovery of New Scale at LHC}
Suppose the LHC sees a classic signal of new physics:
an excess of high transverse energy jets which also have an angular 
distribution that is significantly more isotropic than 
predicted by QCD, an effect that cannot be due to parton distributions 
within the proton.
Suppose further that this measurement corresponds to a scale of new
physics $\Lambda \sim 15$ TeV, 
which is roughly the largest contact interaction that
the LHC could see in the jet channel.  We would have strong 
evidence of new physics, and the angular distribution might begin
to separate between compositeness and other sources of new physics.
But, we would probably not know for certain which source of new 
physics the scale $\Lambda \approx 15$ TeV corresponded too, and we would
need an independent experimental confirmation that quarks were composite.  
If the 
source of new physics were quark compositeness, we would expect to see excited
quarks with mass close to the compositeness scale.
To be safe, we suppose the excited quark mass could be as high as 25 TeV, and 
we want to decide which machine to build to find the excited quark and
confirm that the new physics is quark compositeness. 

\subsection{Discovery of 25 TeV $q^*$ at VLHC}
\begin{figure}[tbh]
\hspace*{-0.25in}
\epsfysize=3.8in
\epsffile{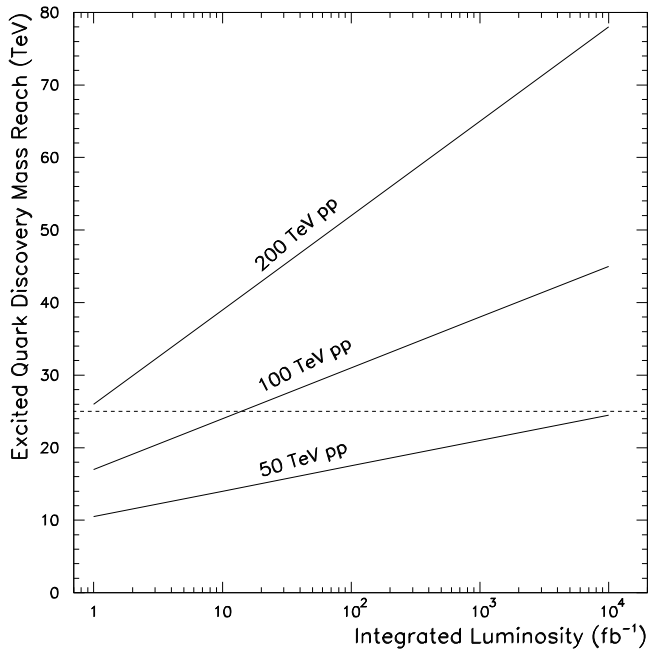}
\caption[]{ 
The discovery mass reach for dijet decays of excited quarks is shown 
as a function of integrated luminosity for a VLHC of energy 50 TeV, 100 
TeV and 200 TeV (solid curves).  The horizontal dashed line is for a 
hypothetical 25 TeV excited quark discussed in the text.}
\label{fig_lum}
\end{figure}

In Fig.~\ref{fig_lum} the horizontal
dashed line at 25 TeV intersects the VLHC excited quark mass reach at 
an integrated luminosity of about $1.3\times10^4$ fb$^{-1}$ for a 50 TeV 
machine, 13 fb$^{-1}$ for a 100 TeV machine, and $0.9$ fb$^{-1}$ for a 200 TeV machine.
Clearly, to find a 25 TeV excited quark, one would build either the 100 TeV
or possible even the 200 TeV machine and quickly accumulate the relatively low
integrated luminosities of 1-10 fb$^{-1}$, rather than build a 50 TeV machine
and have to integrate between 3 and 4 orders of magnitude more luminosity.
Note that the common accelerator wisdom that a factor of 2 in energy is worth
a factor of 10 in luminosity is only roughly right for comparing the 100 TeV
and 200 TeV machines; when comparing the 50 TeV and 100 TeV machines discovery
potential for a 25 TeV excited quark, a factor of 2 in energy is worth a 
factor of $1000$ in luminosity!

\section{Systematics}
In this analysis, we have not included any systematic uncertainties on the
signal, and we have assumed
that the shape and magnitude of the qcd background spectrum is reasonably
approximated by lowest order QCD. We also assumed that the dijet mass 
resolution will be 
roughly 10\% at all hadron colliders, ignoring a long tail to low mass caused 
by radiation.
Adding systematics on the signal and the background will 
likely decrease the mass reach of a real search.
To get a rough idea of the effect of systematics, we examine
the TeV2000 report~\cite{ref_tev2000}, which included systematics in the mass
reach for excited quarks.
Our discovery mass reach for the future 
Tevatron is  about 10\% better than the 95\% CL mass reach quoted in the
TeV2000 report, because ours is for $\sqrt{s}=2.0$ TeV 
instead of 1.8 TeV and because ours does not include systematic uncertainties.
If we increase the mass reach in the TeV2000 report by 10\% to account for
the increase in center of mass energy from $1.8$ to $2.0$ TeV, then the two
results are roughly the same. From this we see that including systematic 
uncertainties would roughly change our $5\sigma$ result to merely a 95\% CL 
result. However, the systematics in the TeV2000 report were likely 
overestimates, because they were based on previous dijet searches for excited 
quarks~\cite{ref_qstar_cdf} in which there was no signal: if a signal is
present the systematic uncertainties will likely be smaller.

\vspace*{-0.5in}
\section{Summary and Conclusions} 
We have estimated the discovery mass reach for excited quarks at future
hadron colliders. The mass reach at the Tevatron is $0.94$ TeV for Run II
(2 fb$^{-1}$) and $1.1$ TeV for TeV33 (30 fb$^{-1}$).  The mass reach at the
LHC is $6.3$ TeV for 100 fb$^{-1}$.  At a VLHC with a center of mass energy
of 50 TeV (200 TeV) the mass reach is 25 TeV (78 TeV) for an 
integrated luminosity of $10^4$ fb$^{-1}$.  However, an excited quark with 
a mass of 25 TeV would be discovered at a hadron collider with $\sqrt{s}=100$
TeV and an integrated luminosity of only 13 fb$^{-1}$: here a factor of 2 
increase in energy from a 50 TeV to a 100 TeV machine is worth a factor of 
$1000$ increase in luminosity at a fixed machine energy of 50 TeV. When the
goal is to discover new physics at high energy scales, even a modest increase 
in machine energy can be more desirable than a large increase in luminosity.

\end{document}